\documentclass[final,5p,times,twocolumn]{elsarticle}
\usepackage{graphicx}
\usepackage{amssymb}
\usepackage{amsmath}
\usepackage{hepunits} 
\usepackage{tabulary}

\usepackage{ifpdf}
\ifpdf
\usepackage{epstopdf}
\usepackage[usenames,dvipsnames]{color}
\usepackage[pdftex,bookmarks=true,hypertexnames=false]{hyperref}
\hypersetup{
  pdfauthor = {Jing Liu},
  pdftitle = {Supernova neutrinos coherently scattered in XMASS},
  pdfsubject = {Supernova neutrinos coherently scattered in XMASS},
  pdfkeywords = {neutrino, coherent scattering, supernova, XMASS},
}
\else
\usepackage[usenames,dvips]{color}
\usepackage[ps2pdf]{hyperref}
\fi
\usepackage{breakurl} 

\biboptions{square,comma,sort&compress}

\journal{Astroparticle Physics}

\begin{document}

\begin{frontmatter}

\title{Detectability of galactic supernova neutrinos coherently scattered on xenon 
nuclei in XMASS}

\author[ICRR,IPMU]{K.~Abe}
\author[ICRR,IPMU]{K.~Hiraide}
\author[ICRR,IPMU]{K.~Ichimura}
\author[ICRR,IPMU]{Y.~Kishimoto}
\author[ICRR,IPMU]{K.~Kobayashi}
\author[ICRR]{M.~Kobayashi}
\author[ICRR,IPMU]{S.~Moriyama}
\author[ICRR]{K.~Nakagawa}
\author[ICRR,IPMU]{M.~Nakahata}
\author[ICRR]{T.~Norita}
\author[ICRR,IPMU]{H.~Ogawa}
\author[ICRR,IPMU]{H.~Sekiya}
\author[ICRR]{O.~Takachio}
\author[ICRR,IPMU]{A.~Takeda}
\author[ICRR,IPMU]{M.~Yamashita}
\author[ICRR,IPMU]{B.~S.~Yang}

\author[IBS]{N.~Y.~Kim}
\author[IBS]{Y.~D.~Kim}

\author[Gifu,TASAKA]{S.~Tasaka}

\author[IPMU,LIU]{J.~Liu}
\author[IPMU]{K.~Martens}
\author[IPMU]{Y.~Suzuki}

\author[Kobe]{R.~Fujita}
\author[Kobe]{K.~Hosokawa}
\author[Kobe]{K.~Miuchi}
\author[Kobe]{N.~Oka}
\author[Kobe]{Y.~Onishi}
\author[Kobe,IPMU]{Y.~Takeuchi}

\author[KRISS,IBS]{Y.~H.~Kim}
\author[KRISS]{J.~S.~Lee}
\author[KRISS]{K.~B.~Lee}
\author[KRISS]{M.~K.~Lee}

\author[Miyagi]{Y.~Fukuda}

\author[STE,KMI]{Y.~Itow}
\author[STE]{R.~Kegasa}
\author[STE]{K.~Kobayashi}
\author[STE]{K.~Masuda}
\author[STE]{H.~Takiya}
\author[STE]{H.~Uchida}

\author[Tokai1]{K.~Nishijima}

\author[YNU1]{K.~Fujii}
\author[YNU1]{I.~Murayama}
\author[YNU1]{S.~Nakamura}

\address{\rm\normalsize XMASS Collaboration$^*$}
\cortext[cor1]{{\it E-mail address:} xmass.publications4@km.icrr.u-tokyo.ac.jp .}

\address[ICRR]{Kamioka Observatory, Institute for Cosmic Ray Research, University of Tokyo, Higashi-Mozumi, Kamioka, Hida, Gifu, 506-1205, Japan}
\address[IBS]{Center of Underground Physics, Institute for Basic Science, 70 Yuseong-daero 1689-gil, Yuseong-gu, Daejeon, 305-811, South Korea}
\address[Gifu]{Information and multimedia center, Gifu University, Gifu 501-1193, Japan}
\address[IPMU]{Kavli Institute for the Physics and Mathematics of the Universe (WPI), University of Tokyo, Kashiwa, Chiba, 277-8582, Japan}
\address[KMI]{Kobayashi-Maskawa Institute for the Origin of Particles and the Universe, Nagoya University, Furo-cho, Chikusa-ku, Nagoya, Aichi, 464-8602, Japan}
\address[Kobe]{Department of Physics, Kobe University, Kobe, Hyogo 657-8501, Japan}
\address[KRISS]{Korea Research Institute of Standards and Science, Daejeon 305-340, South Korea}
\address[Miyagi]{Department of Physics, Miyagi University of Education, Sendai, Miyagi 980-0845, Japan}
\address[STE]{Institute for Space-Earth Environmental Research, Nagoya University, Nagoya, Aichi 464-8602, Japan}
\address[Tokai1]{Department of Physics, Tokai University, Hiratsuka, Kanagawa 259-1292, Japan}
\address[YNU1]{Department of Physics, Faculty of Engineering, Yokohama National University, Yokohama, Kanagawa 240-8501, Japan}

\fntext[TASAKA]{Now at Kamioka Observatory, Institute for Cosmic Ray Research, University of Tokyo, Higashi-Mozumi, Kamioka, Hida, Gifu, 506-1205, Japan.}
\fntext[LIU]{Now at Department of Physics, University of South Dakota, Vermillion, SD 57069, USA.}

\begin{abstract}
  The coherent elastic neutrino-nucleus scattering (CEvNS) plays a crucial role 
  at the final evolution of stars.  The detection of it would be of importance 
  in astroparticle physics.  Among all available neutrino sources, galactic 
  supernovae give the highest neutrino flux in the {\MeV} range. Among all liquid 
  xenon dark matter experiments, XMASS has the largest sensitive volume and 
  light yield.  The possibility to detect galactic supernova via the 
  CEvNS-process on xenon nuclei in the current XMASS detector was investigated.  
  The total number of events integrated in about 18 seconds after the explosion 
  of a supernova 10~kpc away from the Earth was expected to be from 3.5 to 
  21.1, depending on the supernova model used to predict the neutrino flux, 
  while the number of background events in the same time window was measured to 
  be negligible. All lead to very high possibility to detect CEvNS 
  experimentally for the first time utilizing the combination of galactic 
  supernovae and the XMASS detector. In case of a supernova explosion as close 
  as Betelgeuse, the total observable events can be more than $\sim 10^4$, 
  making it possible to distinguish different supernova models by examining the 
  evolution of neutrino event rate in XMASS.
\end{abstract}


\end{frontmatter}


\section{Introduction}
\label{s:intro}
The neutral current interaction of neutrinos with nuclei,
\begin{equation}
  \nu_{\mu}/\bar{\nu}_{\mu} + \text{nucleus} \rightarrow
  \nu_{\mu}/\bar{\nu}_{\mu} + \text{hadrons},
  \label{e:hen}
\end{equation}
was observed at the Gargamelle bubble chamber experiment at CERN in 
1973~\cite{garg73}. The energy of incident neutrinos was in the order of a few 
\GeV. Freedman pointed out one year later~\cite{freedman74, freedman77} that a 
neutrino with an energy in the order of \MeV could interact with all nucleons 
in a nucleus coherently,
\begin{equation}
  \nu + \text{nucleus} \rightarrow \nu + \text{nucleus},
  \label{e:len}
\end{equation}
resulting in a large cross section, approximately proportional to the square of 
the number of neutrons in the target nucleus. Given such a large cross section, 
however, it has not been observed yet, primarily because the only observable of 
this interaction is the recoiled nucleus with its kinetic energy in the order 
of \keV.

Although not yet observed, the coherent scattering has been believed to be the 
main mechanism for neutrinos to be trapped in the core of a 
supernova~\cite{janka07}. It has been proposed as a method to probe 
non-standard neutrino interactions with quarks, extra heavy neutral gauge 
bosons~\cite{barranco05} and the neutron part of nuclear form 
factors~\cite{patton12}.  A detector utilizing the coherent scattering was also 
proposed by Drukier and Stodolsky in 1984~\cite{drukier84} to detect neutrinos 
from spallation sources, reactors, supernovae, the Sun and the Earth, and by 
Goodman and Witten in 1985~\cite{goodman85} to detect some dark matter 
candidates.  Among all available neutrino sources, galactic supernovae give the 
highest neutrino flux in the \MeV range.

The coherent scattering has been listed as the primary physics goal of many 
experimental proposals, such as CoGeNT~\cite{barbeau07}, TEXONO~\cite{wong08}, 
NOSTOS~\cite{gio08}, RED~\cite{red12,red16}, CosI~\cite{cosi15}, 
COHERENT~\cite{coherent15}, CENNS~\cite{cenns14} and CONNIE~\cite{connie16} 
{\it etc}. In addition, experiments for low-energy solar neutrino, dark matter 
and neutrinoless double beta-decay have the potential to detect galactic 
supernova neutrinos coherently scattered on nuclei.  Horowitz made a 
comprehensive comparison between different approaches~\cite{horowitz03}. He 
pointed out that the choice of the target nuclei involved a trade-off of many 
considerations.  For example, heavy elements are preferred because of the large 
cross section, while light elements are preferred since they get more recoil 
energy, which relaxes the requirement on low energy threshold. And most 
importantly, the amount of target material should be as large as possible. 
Among all existing experiments, liquid xenon dark matter experiments seem to be 
the most practical choice for this purpose at this moment, given their large 
target masses, sufficiently low background and energy thresholds.

Most xenon based dark matter experiments utilize dual-phase time-projection 
chambers. Their potential to detect supernova neutrinos through the CEvNS 
channel is discussed in a recent paper~\cite{lang16} and references there in. 
The XMASS detector~\cite{xmdet} located in the Kamioka Underground Observatory 
in Japan is a single-phase liquid-xenon scintillation detector. It contains the 
largest amount of liquid xenon and features the highest light yield among all 
running liquid-xenon dark-matter experiments.  The high light yield ensures a 
sufficiently low energy threshold, while the background level around the 
threshold was measured to be negligible~\cite{xmlight} in a 18~second time 
window, the typical time scale of a supernova neutrino burst. All make XMASS a 
promising experiment to detect supernova neutrinos through the coherent 
scattering channel. Neutrino-electron neutral current scatterings ($\nu + e^{-} 
\rightarrow \nu + e^{-}$) and neutrino-nucleus quasi-elastic scatterings are 
other possible observation channels. However, their cross sections are orders 
of magnitude smaller than that of coherent scatterings~\cite{marciano03,
  chasioti09, ydrefors12, divarixe132, almosly13, divari13}, and will not be 
  discussed in this work.

Several detectors in the Kamioka Underground Observatory are capable of 
detecting supernova neutrinos along with many others in the 
world~\cite{scholberg12, mirizzi15}. The water \v{C}erenkov detector 
Super-Kamiokande can detect supernova neutrinos dominantly through the 
$\bar{\nu_{e}} + p \rightarrow e^+ + n$ channel~\cite{totani98, ikeda07}. 
Utilizing the neutral current interaction, XMASS is sensitive to all flavors of 
neutrinos.  Another experiment, KamLAND, is also sensitive to all flavors 
through the neutrinos-proton elastic scattering~\cite{beacom02, dasgupta11} and 
excitation of carbon nuclei by neutrinos~\cite{gando12}. However, no 
information on the coherent scattering is given by KamLAND.  The three
experiments in Kamioka cover each other's dead time, are sensitive to different
neutrino interactions and may provide comprehensive understanding of the
supernova neutrino burst in case of a simultaneous observation.

The possibility to detect galactic supernova neutrinos coherently scattered 
with xenon nuclei in XMASS is calculated in this work. Since XMASS is a running 
detector with most of its properties having been studied 
systematically~\cite{xmdet, xmlight}, the uncertainty of the estimation in the 
detection is minimized.  However, the precision of such an estimation still 
suffers from the uncertainty in the theoretical prediction of supernova neutrino 
flux, as demonstrated recently by Chakraborty \textit{et al.}~\cite{chak13}.

In order to have a comprehensive understanding of the possible variation in the 
event rate predicted by various supernova models, the numeric database of 
supernova neutrino emission provided by Nakazato \textit{et 
al.}~\cite{nakazato13} is used to calculate the coherent scattering event-rate 
in XMASS.  Numeric results of a wide range of progenitors are provided 
including a black-hole-forming case.  The canonical Livermore supernova 
model~\cite{totani98} is also used to calculate the event rate, the result of 
which can be used as a reference when compared to other estimations.

Two possible locations of galactic supernovae are assumed. One is 10~kpc away 
from the Earth, roughly at the center of the Milky Way. The other is 196~pc 
away from the Earth where Betelgeuse locates.

\section{Coherent elastic neutrino-nucleus scattering}
\label{s:xs}
The differential cross section of the coherent scattering as a function of 
neutrino energy $E_{\nu}$ and nuclear recoil energy $E_{\text{nr}}$ is taken 
from Ref.~\cite{barranco05},
\begin{equation}
    \frac{\mathrm{d}\sigma}{\mathrm{d}E_{\text{nr}}}(E_{\nu},E_{\text{nr}}) = 
    \frac{G^{2}_\text{F}M}{2\pi} G_\text{V}^2 
    \left[\vphantom{\frac{1}{2}}\right. 1 + 
    \left(1-\frac{E_{\text{nr}}}{E_{\nu}}\right)^2 - 
  \frac{ME_{\text{nr}}}{E^2_{\nu}} \left.\vphantom{\frac{1}{2}}\right],
  \label{e:xs}
\end{equation}
where $G_\text{F}$ is the Fermi constant, $M$ is target nuclear mass, and
\begin{eqnarray}
  G_\text{V} &=& [(\frac{1}{2}-2\sin^2\theta_W)Z -\frac{1}{2}N]F(q^2),
  \label{e:gv}
\end{eqnarray}
excluding non-standard neutrino interaction terms and neglecting the radiative 
corrections presented in Ref.~\cite{barranco05}. The axial vector current leads 
to a small incoherent contribution to the total neutral current cross section 
and is ignored. The value of $\sin^2\theta_W$ ($\theta_W$ is the weak mixing 
angle) is 0.23, taken from the Review of Particle Physics~\cite{pdg}.  $Z$ 
and $N$ are the numbers of protons and neutrons in the nucleus, respectively.  
According to the definition in Ref.~\cite{freedman77}, the nuclear form factor 
$F(q^2)$ is the integral of the relative phase of the incident neutrino 
scattered by the nucleon at position $\mathbf{r}$:
\begin{equation}
  F(q^2) = \int \mathrm{d}\mathbf{r}
  e^{i\mathbf{q}\cdot\mathbf{r}} \rho(\mathbf{r}),
  \label{e:f}
\end{equation}
where $\rho(\mathbf{r})$ is the spatial density distribution of nucleons, 
normalized so that $\int\mathrm{d}\mathbf{r}\rho(\mathbf{r})=1$. 
Helm proposed to reform it as $\rho(\mathbf{r}) = \int \mathrm{d}\mathbf{r}' 
\rho_0(\mathbf{r}') \rho_1(\mathbf{r} - \mathbf{r}')$~\cite{helm56}, where 
$\rho_0$ represents a constant density inside a sphere with radius $r_0$, and 
$\rho_1$ a surface with thickness $s$. The form factor can then be expressed 
as~\cite{engel91}
\begin{equation}
  F(q^2) = \frac{3 j_1(qr_0)}{qr_0} e^{-\frac{1}{2}(qs)^2},
  \label{e:fj}
\end{equation}
where $j_1(qr_0) = [\sin(qr_0) - qr_0\cos(qr_0)]/(qr_0)^2$ is the spheric 
Bessel function of the first order. The relation between nuclear radius $r_n$ 
and $r_0$~\cite{helm56, engel91, lewin96} is $r_0^2 = r_n^2 - 5 s^2$. Values 
of $r_n$ are taken from Ref.~\cite{fricke95} and listed in Table~\ref{t:xe}. 
The value of $s$ is taken as 1~fm~\cite{engel91}.

\begin{table}[htbp]
  \centering
  \caption{Properties of natural xenon isotopes used in calculation.}
  \label{t:xe}
  \begin{tabulary}{\linewidth}{CCCC}\hline
    Natural    &Nuclear     & Natural    & Nuclear\\
    xenon      &mass        & abundance  & radius \\
    isotope    &(\GeV/c$^2$)& (atomic \%)& (fm)   \\\hline
    $^{128}$Xe &119.1147    & 1.92       & 4.776  \\
    $^{129}$Xe &120.0474    &26.44       & 4.776  \\
    $^{130}$Xe &120.0777    & 4.08       & 4.783  \\
    $^{131}$Xe &121.9107    &21.18       & 4.781  \\
    $^{132}$Xe &122.8413    &26.89       & 4.787  \\
    $^{134}$Xe &124.7055    &10.44       & 4.792  \\
    $^{136}$Xe &126.5702    & 8.87       & 4.799  \\\hline
  \end{tabulary}
\end{table}

\section{Core-collapse supernovae}
\label{s:mod}
\begin{figure}
  \includegraphics[width=\linewidth]{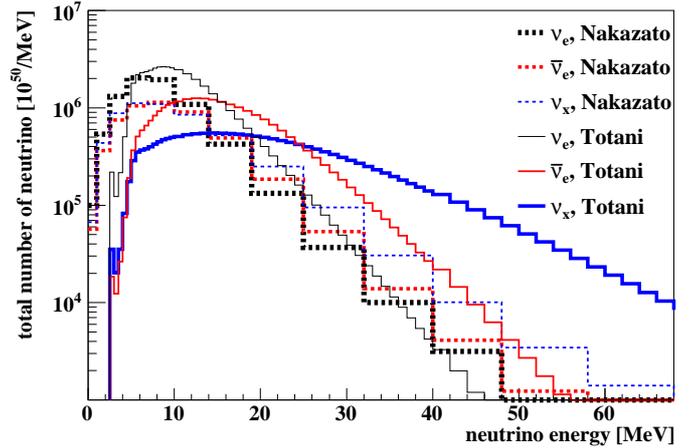}
  \caption{Supernova neutrino energy-spectra integrated from the core collapse 
    to about 18~seconds.  Solid lines are the numeric results from Totani 
    \textit{et al.}~\cite{totani98}. Dashed lines are the numeric results from 
    Nakazato \textit{et al.}~\cite{nakazato13}. The parameters used to generate 
    the spectra from Nakazato model are progenitor mass ($M_p = 
    20$~M$_{\bigodot}$), metalicity ($Z = 0.02$) and shock wave revival time 
    ($t_{\text{rev}} = 200$~ms). $\nu_x$ represents neutrino flavors other than 
    $\nu_e$ and $\bar{\nu}_e$.
  } \label{f:snve}
\end{figure}

Stars heavier than $~8$~M$_{\bigodot}$ end their lives as core-collapse
supernovae.  It is commonly believed that the shock wave losses its kinetic
energy when propagating outward and stalls before blowing off the stellar
envelope. Several mechanisms causing the shock wave to revive have been
proposed~\cite{wilson82, bethe85, bethe90, burrows06, fischer11, kotake06}.
Different models predict different shock wave revival times. Generally
speaking, the later the revival, the more neutrinos are emitted because more
material falls on to the acretion shock. This causes an uncertainty in the
expected number of events observed in a detector. Nakazato \textit{et
al.}~\cite{nakazato13} proposed a simple method to manually combine their one
dimensional simulations before and after shock wave revive. The revival time
$t_{\text{rev}}$ is used as a parameter related to the yet unknown explosion
mechanism.  The number luminosity and energy spectrum of neutrinos as a
function of time are provided by them in a publicly accessible
database~\cite{db}. The results corresponding to $t_{\text{rev}} = 100, 200$
and 300~ms are provided in the current database. The influence of
$t_{\text{rev}}$ on the observed energy spectra and event rates can be
investigated using those results.  Other parameters that can be investigated
using this database include the masses of supernova progenitors, $M_{p}$,
($M_{p} = 13, 20, 30$ and 50~M$_{\bigodot}$ are provided) and the metalicity,
$Z$, of the galaxy where those progenitors are located ($Z = 0.02$ and its 1/5
are provided). They are all used in this paper.  The simulation result from
Totani \textit{et al.} published in 1998~\cite{totani98} has been widely used
in previous calculations. It is also used in this paper to provide a reference
for comparison.  Figure~\ref{f:snve} shows the neutrino energy spectra 
integrated from the core collapse till about 18~seconds later, provided by 
Totani \textit{et al.} and Nakazato \textit{et al}.  The parameters used to 
generate the spectra from Nakazato model are $M_p = 20$~M$_{\bigodot}$, $Z = 
0.02$ and $t_{\text{rev}} = 200$~ms. The total energy carried by neutrinos in 
this model is $1.92 \times 10^{53}$~erg. The average energies of different 
neutrino flavors given by this model are 9.32~MeV, 11.1~MeV and 11.9~MeV for 
$\nu_e, \bar{\nu}_e$ and $\nu_x$, respectively. The energy release as a 
function of the three input parameters is summarized in Table~1. in 
reference~\cite{nakazato13}.

\section{XMASS detector}
\label{s:det}
The key factors for a successful detection of galactic supernova neutrinos 
coherently scattered in a detector include large target mass, low energy 
threshold and low background. All of them are fulfilled in the current XMASS 
detector~\cite{xmdet}.  It is a liquid-xenon scintillator-detector, and the 
scintillation light is collected by 642 photomultiplier tubes (PMTs) mounted on 
a pentakis-dodecahedral support structure with a diameter of about 80~cm.  The 
active region contains $M_{\text{det}}=832$~kg of liquid xenon, the largest 
among all liquid xenon dark matter experiments. The photocathode coverage of 
the inner surface of the detector is 62.4\%.  A global trigger is generated if 
the number of hit PMTs within a 200~ns window is above three. The detector is 
located underground in the Kamioka Observatory at a depth of 
2700~meter-water-equivalent. The cosmic ray induced background is sufficiently 
suppressed.  To shield the scintillator volume from external gammas, neutrons, 
and muon-induced backgrounds, the copper vessel is placed at the center of a 
cylindrical tank filled with pure water with a diameter of 10~m and a height of 
11~m. This volume is viewed by 72 Hamamatsu R3600 20-inch PMTs to provide both 
an active muon veto and passive shielding against these backgrounds.  The 
background level around the threshold was measured in the commissioning 
runs~\cite{xmlight}.  XMASS started the physics run in November 2013 after the 
detector refurbishment and the background rate is further 
reduced~\cite{hiraide2015} to be negligible in a 18~second time window, the 
typical time scale of a supernova neutrino burst.  This makes it possible to 
utilize all sensitive volume of XMASS for supernova neutrino detection.

\section{Energy spectra of supernova neutrino events}
\label{s:spec}
The differential event rate of supernova neutrinos in the liquid xenon target 
in XMASS as a function of the true nuclear recoil energy $E_{\text{nr}}$ can be 
expressed as:
\begin{equation}
  \frac{\mathrm{d}R_0}{\mathrm{d}E_{\text{nr}}}(E_{\text{nr}}) =
  \frac{M_{\text{det}} N_{A}}{A (4\pi d^2)} \sum_{i=\nu_{e}, \bar{\nu}_{e}, 
  \nu_x}
  \int^{\infty}_{E_{\text{min}}}
  \frac{\mathrm{d}\sigma}{\mathrm{d}E_{\text{nr}}}(E_{\nu}, E_{\text{nr}})
  f_{i}(E_{\nu})\mathrm{d}E_{\nu},
  \label{e:dr0}
\end{equation}
where $N_A$ is the Avogadro's number, $A$ is the averaged atomic mass of 
natural xenon and $d$ is the distance between the supernova and the detector, 
$E_{\text{min}} = (E_{\text{nr}} + \sqrt{E_{\text{nr}}^2 + 2 M E_{\text{nr}}})/2$ 
is the minimum energy a neutrino must have in order to give to the nucleus a 
recoil energy $E_{\text{nr}}$, and $f_{i}(E_{\nu})$ is the neutrino energy 
spectra shown in Figure~\ref{f:snve}.

\begin{figure}
  \includegraphics[width=\linewidth]{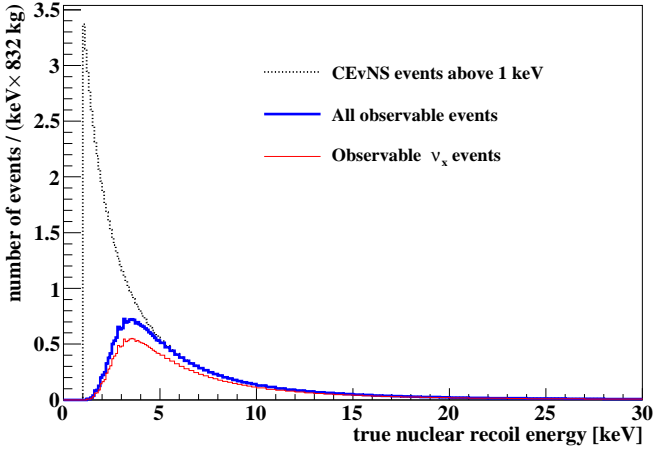}
  \caption{Energy spectrum as a function of true xenon nuclear recoil energy. 
    The upper most and middle curves are the energy spectra with and without 
    taking into account the detection efficiency, respectively. The lower most 
    curve shows the contribution from $\nu_{x}$ only. The upper line is 
    calculated above 1~\keV.  The supernova model used here is the one from 
    Nakazato \textit{et al.} with $M_p = 20$~M$_{\bigodot}, Z = 0.02$ and 
    $t_{\text{rev}} = 200$~ms. This specific model predicts neither most nor 
    least neutrino flux, hence is chosen to create the plot in order to avoid 
    any visual bias.  The distance of the supernova from the Earth is assumed 
  to be 10~kpc.} 
    \label{f:er}
\end{figure}

The upper most curve in Figure~\ref{f:er} shows the true recoil energy spectrum 
calculated with Equation~\ref{e:dr0} above 1~\keV.  Nuclear recoils below 
{1~\keV} create less than 1 photoelectron assuming standard liquid xenon 
scintillation efficiency~\cite{leff} and the light yield recorded in 
XMASS~\cite{xmdet}, hence are ignored.

The full XMASS Monte Carlo simulation is used to estimate the detection 
efficiency $\varepsilon(E_{\text{nr}})$. The upper most curve in 
Figure~\ref{f:er} is used to sample the recoil energies of xenon nuclei as 
input for this Geant4-based simulation.  The quenching of nuclear recoil energy 
in the scintillation process, the optical properties of liquid xenon, copper 
and PMTs, the quantum efficiency of PMTs and the electronic smearing of the 
number of photoelectrons are all implemented~\cite{xmdet} in addition to the 
tracking process provided by Geant4. A PMT with the number of photoelectrons 
above 0.25 is recorded as a hit. The total number of hits, $N_{\text{hits}}$, 
is recorded for each simulated event. The detection efficiency 
$\varepsilon(E_{\text{nr}})$ is defined as the fraction of events with 
$N_{\text{hits}}>3$ at a certain recoil energy $E_{\text{nr}}$.  The realistic 
recoil energy spectrum is then
\begin{equation}
  \frac{\mathrm{d}R}{\mathrm{d}E_{\text{nr}}}(E_{\text{nr}}) =
  \varepsilon(E_{\text{nr}}) \times 
  \frac{\mathrm{d}R_0}{\mathrm{d}E_{\text{nr}}}(E_{\text{nr}})
  \label{e:dr}
\end{equation}
as shown in the middle curve in Figure~\ref{f:er}.

\begin{figure}
  \includegraphics[width=\linewidth]{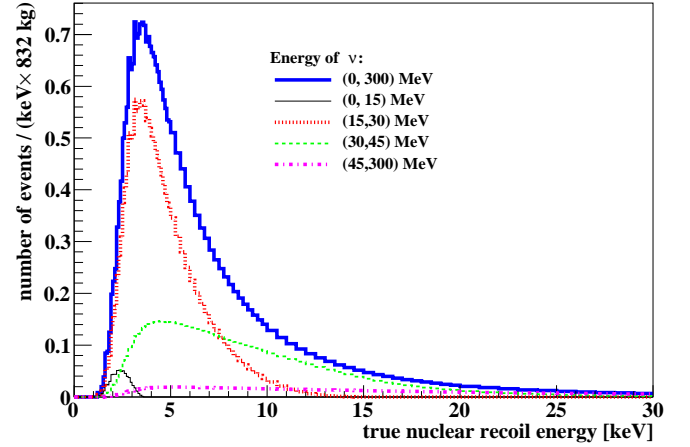}
  \caption{Sensitivity to various supernova neutrino energy regions. The upper 
  most curve is exactly the same as the middle curve in figure~\ref{f:er}.}
  \label{f:ev}
\end{figure}

The lower most curve in Figure~\ref{f:er} shows the contribution to the 
observable energy spectrum from $\nu_x$ only. Clearly, XMASS detects mostly 
$\nu_x$. The upper most curve in Figure~\ref{f:ev} is exactly the same as the 
middle curve in Figure~\ref{f:er}. The lower curves in Figure~\ref{f:ev} show 
contributions from neutrinos in various energy regions. Due to the threshold 
effect, XMASS is mostly sensitive to neutrinos above $\sim 15$~MeV from the 
tail parts of the supernova neutrino spectra shown in Figure~\ref{f:snve}.

The total number of observable events, $N_{\text{obs}}$, can be obtained by 
integrating the realistic energy spectrum:
\begin{equation}
  N_{\text{obs}} = \int 
  \frac{\mathrm{d}R}{\mathrm{d}E_{\text{nr}}}(E_{\text{nr}})
  \mathrm{d}E_{\text{nr}}
  \label{e:nobs}
\end{equation}
Practically, it is enough to integrate over $E_{\text{nr}} = 1$-{50~\keV} as seen 
in the middle curve of Figure \ref{f:er}.  The values of $N_{\text{obs}}$ from 
different supernova models are listed in Table~\ref{t:nobs}. Two distances are 
chosen for comparison, $d = 10$~kpc is roughly the distance from the center of 
the Milky Way to the Earth, $d = 196$~pc is the distance from Betelgeuse to the 
Earth. The number of observable events predicted by most of the Nakazato models 
are significantly less than that predicted by the Livermore model. However, one 
Nakazato model, which forms a black-hole, predicts similar number of observable 
events as the Livermore model. This points out the possibility to detect failed 
supernovae with no optical signal. In case of a supernova as close as 
Betelgeuse, all the models predict a definitely possible observation.

\begin{table} [htbp]
  \centering
  \caption{Number of observable supernova events in XMASS. The weakest Nakazato 
    model is the one with $M_{p} = 20$~M$_{\bigodot}$, $Z = 0.02$ and 
    $t_{\text{rev}} = 100$~ms.  The brightest Nakazato model is the one with 
    $M_{p} = 30$~M$_{\bigodot}$, $Z = 0.02$ and $t_\text{rev} = 300$~ms. The 
    black-hole-forming model is the one with $M_{p} = 30$~M$_{\bigodot}, Z = 
    0.004$. Neutrino energy spectra used in the calculation are all integrated 
    from core collapse till about 18~seconds later.
  } \label{t:nobs}
  \begin{tabulary}{\linewidth}{LCC}\hline
    Supernova model & $d = 10$~kpc &  $d = 196$~pc \\\hline
    Livermore & 15.2 & $3.9\times10^{4}$ \\
    Nakazato (weakest) & 3.5 & $0.9\times10^{4}$ \\
    Nakazato (brightest) & 8.7 & $2.3\times10^{4}$ \\
    Nakazato (black hole) & 21.1 & $5.5\times10^{4}$ \\\hline
  \end{tabulary}
\end{table}

The energy of an event in XMASS is estimated by converting the recorded number 
of photoelectrons to \keV{} using a measured relationship between these two.  
Such a relationship is obtained in energy calibrations at various locations in 
the detector using $\gamma$ or $X$-ray sources with different energies as 
detailed in Ref.~\cite{xmdet}. There is less than 10\% difference in the energy 
converted this way from events with the same number of photoelectrons but at 
different locations~\cite{xmmodul}. Due to the fact that the scintillation 
efficiencies of nuclear and electronic recoil events are different~\cite{leff}, 
the energy calibrated this way is called explicitly \emph{electron equivalent} 
energy to avoid ambiguity.  The energy resolution is 36\% at 1~\keV{} (electron 
equivalent), dominated by Poisson statistics~\cite{xmmodul}.

\begin{figure}
  \includegraphics[width=\linewidth]{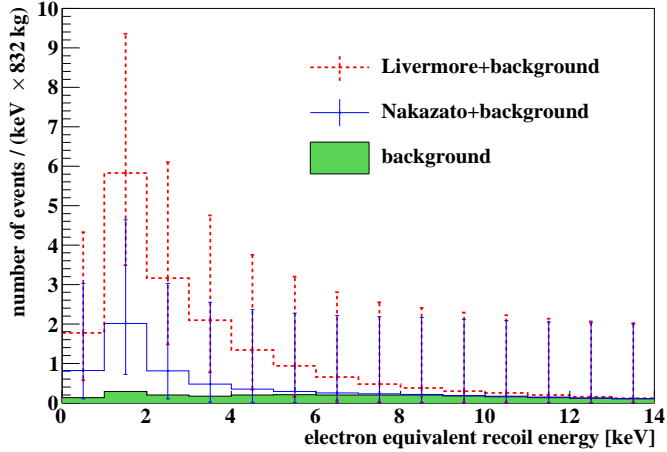}
  \caption{The solid histogram in the middle is the same recoil energy 
    spectrum as the middle curve of Figure~\ref{f:er}, but in the unit of 
    \emph{electron equivalent} instead of the \emph{true nuclear recoil} 
    energy. The dashed histogram on the top is the \emph{electron 
    equivalent} recoil energy spectrum of the Livermore model. Both of them are 
    draw on top of the expected background estimated from XMASS measurements, 
  shown as the filled histogram on the bottom.} \label{f:kevee}
\end{figure}

The solid histogram in the middle of Figure~\ref{f:kevee} shows the same 
recoil energy distribution as the middle curve of Figure~\ref{f:er}, but in the 
unit of \emph{electron equivalent} recoil energy instead of the \emph{true 
nuclear recoil} energy. It is converted from the distribution of number of 
photoelectrons obtained from the full XMASS simulation.  The spectrum is 
plotted on top of the expected background spectrum estimated from XMASS 
measurements, shown as the filled histogram on the bottom of 
Figure~\ref{f:kevee}. The error bars represent Poisson 68\% CL intervals. The 
error bars in the background spectrum are invisibly small.  For comparison, the 
\emph{electron equivalent} recoil energy spectrum of the Livermore model is 
generated the same way and shown as the dashed histogram on the top of 
Figure~\ref{f:kevee}.

\section{Event rate}
\label{s:er}
\begin{figure}
  \includegraphics[width=\linewidth]{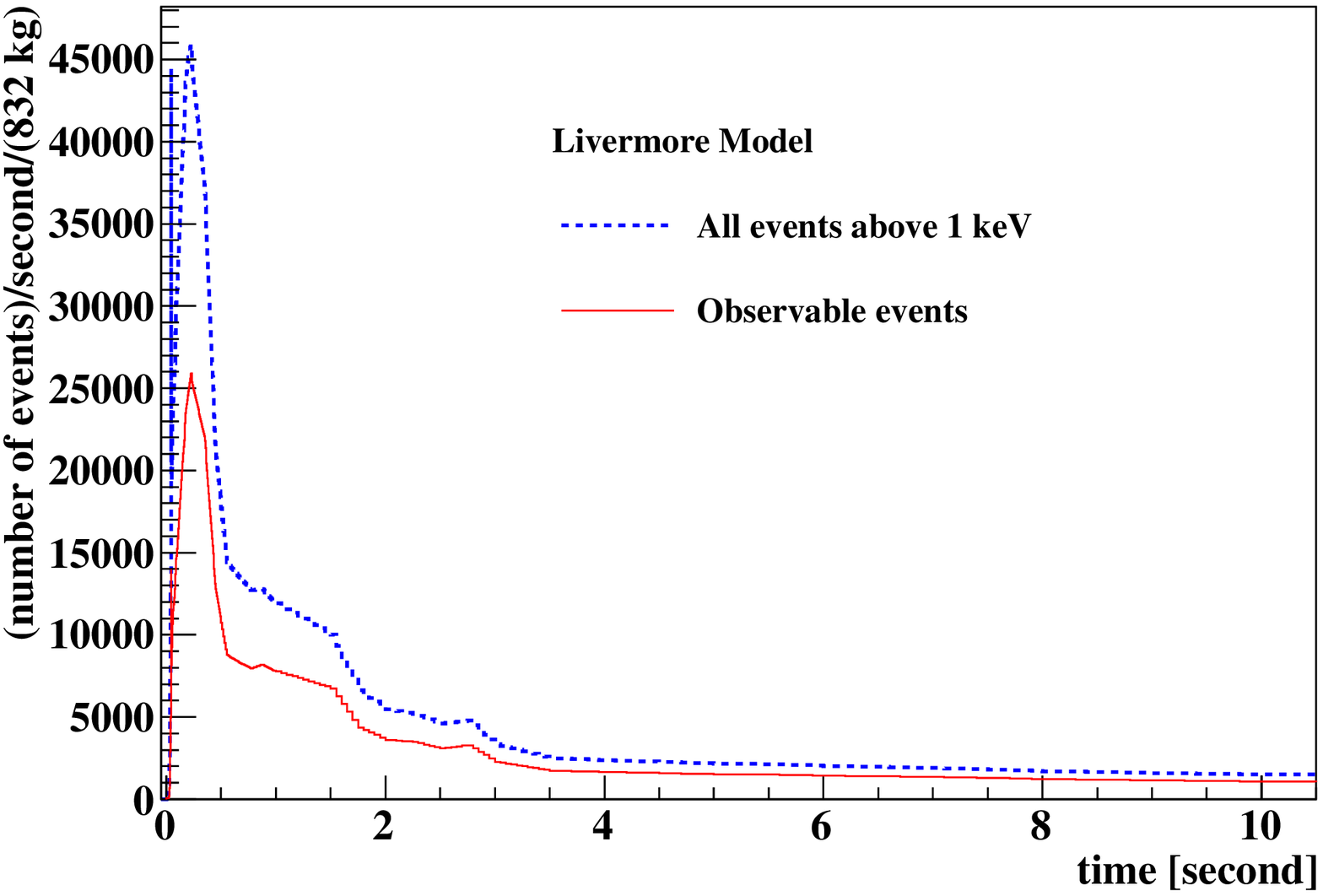}\\
  \includegraphics[width=\linewidth]{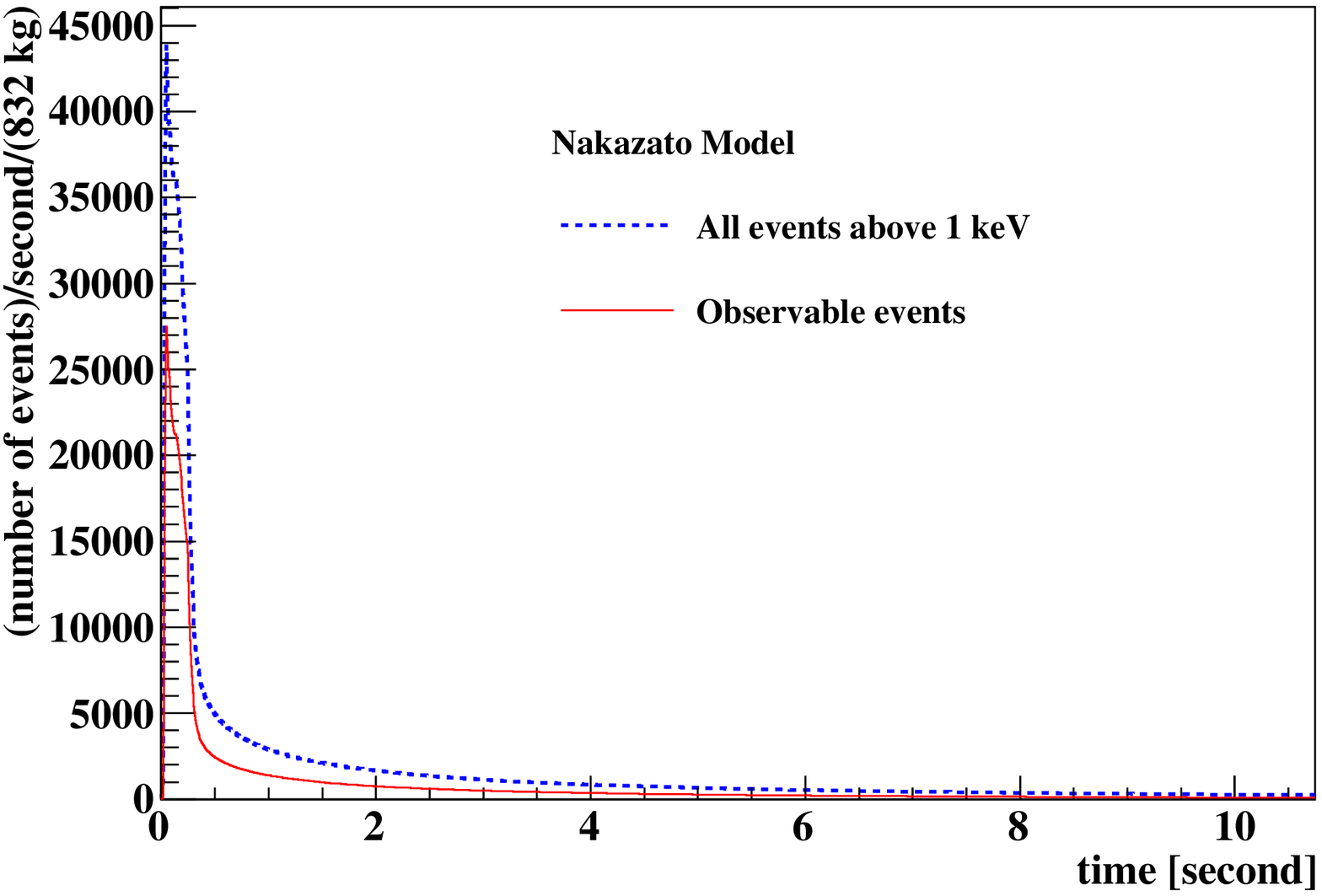}
  \caption{Rate of CEvNS events in XMASS for a supernova 196~pc away from the 
    Earth predicted by the Livermore model (upper plot) and the Nakazato model 
    with $M_{p} = 20$~M$_{\bigodot}$, $Z = 0.02$ and $t_{\text{rev}} = 200$~ms 
    (lower plot). This specific Nakazato model predicts neither most nor lest 
    neutrino flux, hence is chosen to create the plot in order to avoid any 
    visual bias. The upper lines correspond to all the CEvNS events above 1 keV 
    nuclear recoil energy predicted by models; the lower lines corresponds to 
    all events that can be detected in XMASS. About half of the events are 
  detectable.} \label{f:rate}
\end{figure}

As shown in Table~\ref{t:nobs}, the average event rate in XMASS can be as high 
as a few thousand events per second for a supernova as close as Betelgeuse. 
Given such a high rate, it is possible to study in detail the supernova 
explosion mechanism by examining the time evolution of the event rate, since 
the flux and energy of the neutrinos predicted by different models vary in 
different phases of the explosion.  Figure~\ref{f:rate} shows the rate of CEvNS 
events in XMASS in about 18~second for a supernova 196~pc away from the Earth
predicted by the Livermore model and Nakazato model with $M_{p} =
20$~M$_{\bigodot}$, $Z = 0.02$ and $t_{\text{rev}} = 200$~ms, assuming without
any DAQ loss. Different supernova models can be clearly distinguished.

Event rates of other neutrino interactions such as neutrino-electron neutral 
current scatterings and neutrino-nucleus quasi-elastic scatterings are not 
negligible in this case. Possible optimization of XMASS electronic system is 
under investigation to cope with such a high event rate.

\section{Conclusion}
\label{s:conc}
The possibility to detect galactic supernova neutrinos coherently scattered 
with xenon nuclei in XMASS was examined in detail.  The predicted number of
observed events depend on two factors, one is the detection efficiency of the 
detector at low nuclear recoil energy, the other is the neutrino flux predicted 
by the supernova model used for the calculation. The former is estimated using
full XMASS simulation. The latter is estimated by examining all models
available in Nakazato's database~\cite{nakazato13}.  The predicted number of
observable events in XMASS ranges from 3.5 to 21.1 for a supernova in the
center of the Milky Way, while the number of background events in the same time
and energy window is observed to be negligible. It is hence possible for XMASS
to experimentally observe galactic supernova neutrinos coherently scattered on
xenon nuclei for the first time.  In case of a supernova as close as
Betelgeuse, the average event rate is above thousands per second, making it
possible to distinguish between different supernova models by examining the
time evolution of the event rate. Such a detection would provide not only the
experimental evidence of CEvNS, but also comprehensive information about the
supernova explosion mechanism.

\section*{Acknowledgments}
\label{s:ackn}

We gratefully acknowledge the cooperation of the Kamioka Mining and Smelting 
Company. This work was supported by the World Premier International Research 
Center Initiative (WPI Initiative), MEXT, Japan, the Grant-in-Aid for 
Scientific Research, MEXT, Japan, JSPS KAKENHI Grant No. 19GS0204, 26104004, 
26105505 and 26104007 and partially by the National Research Foundation of 
Korea Grant funded by the Korean Government (NRF-2011-220-C00006).





\bibliographystyle{elsarticle-num}
\bibliography{SNvXMASS}







\end{document}